\documentclass[amsfonts,smallextended]{article}

\usepackage{epsfig} 

\usepackage{citesort}

\begin{document}

\title{Universally Coupled Massive Gravity, II: Densitized Tetrad and Cotetrad Theories\footnote{Forthcoming in \emph{General Relativity and Gravitation}}}

\author{J. Brian Pitts\footnote{Department of Physics, Department of Philosophy, and Reilly Center for Science, Technology and Values, University of Notre Dame, Notre Dame, Indiana 46556 USA, (574) 904-1177, jpitts@nd.edu	}}
 	

\maketitle


\begin{abstract}

   Einstein's equations in a tetrad formulation are derived from  a linear theory in flat spacetime with an asymmetric  potential using free field gauge invariance, local Lorentz invariance and universal coupling.  The gravitational potential can be either covariant or contravariant and of almost any density weight.  These results are adapted to produce universally coupled massive variants of Einstein's equations, yielding two one-parameter families of distinct theories with spin 2 and spin 0.  The theories derived, upon fixing the local Lorentz gauge freedom, are seen to be a  subset of those found by   Ogievetsky and  Polubarinov some time ago using a spin limitation principle.  In view of the stability question for massive gravities, the proven non-necessity of positive energy for stability in applied mathematics in some contexts is recalled.  Massive tetrad gravities permit the mass of the spin 0 to be heavier than that of the spin 2, as well as lighter than or equal to it, and so provide phenomenological flexibility that might be of astrophysical or cosmological use.

\end{abstract}

\vspace{.25in}
keywords: massive graviton, tetrad, universal coupling, positive energy, nonlinear stability, tensor density


\section{Introduction}

The project of deriving a relativistic gravitational theory using considerations such as an analogy to Maxwellian electromagnetism, the universal coupling of the gravitational field to a combined gravity-matter energy-momentum complex, and the requirement that the gravitational field equations alone (without the matter equations) entail energy-momentum conservation  was a part of Einstein's search for an adequate theory of gravity in 1913-15 \cite{EinsteinPapers8,EinsteinTrans4,Janssen,JanssenRenn}. 
This history was subsequently downplayed by Einstein \cite{JanssenRenn,Feynman}, and the above ideas later came to be associated with the (allegedly  non-Einsteinian) field-theoretic approach to gravitation more commonly associated with particle physics, where it was brought to successful completion in the 1950s \cite{Kraichnan,Gupta}.  

For some time it was believed that there was a unique massive generalization of General Relativity satisfying universal coupling \cite{FMS,DeserMass}. This theory has the same mass for the spin $0$ as for the spin $2$, as well as a specific nonlinear algebraic  self-interaction such that the DeDonder-Fock harmonic condition is satisfied by the effective curved metric when Cartesian coordinates are used for the flat background metric.

 A few years ago it was found that there are, at least, two one-parameter families of universally coupled massive theories \cite{MassiveGravity1}. These involve either a covariant or contravariant symmetric tensor potential of almost any density weight; the resulting theories involve mass terms conveniently presented in terms of correspondingly densitized curved metric or inverse metric.   All of these theories are included in the 2-parameter family of theories found by Ogievetsky and Polubarinov in terms of their spin limitation principle \cite{OP,OPMassive2},  which imposed  axiomatically an auxiliary condition to eliminate the spin $1$ and one spin $0$ degrees of freedom.

 An important but often neglected aspect of deriving Einstein's equations as the field equations for a self-interacting massless spin $2$ field theory is the inclusion of spin $\frac{1}{2}$ matter, that is, fermions.  This problem was addressed using in effect an orthonormal tetrad formalism in an unjustly neglected paper by Shirafuji  \cite{Shirafuji}. 
Going back to Weyl in 1929 \cite{WeylGravitationElectron,CartanSpinor}, it  is  often claimed  that coupling of spinors to the curved space-times implied by General Relativity requires the introduction of an orthonormal tetrad.  While that claim was proven false in the mid-1960s \cite{OP,OPspinor,GatesGrisaruRocekSiegel},  nonetheless in some contexts it is \emph{convenient}  to represent the gravitational potential using an orthonormal tetrad, especially (but not only) if spinors are present, in order to avoid technically demanding nonlinearities. 
 The interesting question of including spinor matter is set aside for another occasion, however, in favor of deriving mass terms that can accompany the Einstein tensor.

Using a tetrad formalism,  I will  exhibit two one-parameter families of massive gravities which have not previously been derived from universal coupling.   Einstein's equations have been derived using universal coupling with a tetrad and an \emph{independent} connection \cite{DeserFermion,DeserSupergravity}.  While this paper  draws on themes from those papers, it aims at deriving Einstein's theory and then massive generalizations, treating the  connection implicitly as some function of the tetrad.  It seems to me that the present derivation, which does not aim to arrive at supergravity or even gravity with spinors, has the  advantage of requiring little knowledge of the answer beforehand, as well as the simplicity of fewer field variables.  It is therefore quite useful in the context of discovery, not just the context of justification. 


\section{Stability or Empirical Adequacy Questions for Massive Gravity}

From the early 1970s until roughly 2000 (as dark energy was becoming accepted), the conventional wisdom about (Lorentz-invariant) massive variants of General Relativity was that they came in two varieties, which succumb to different problems \cite{DeserMass,vDVmass1,Zakharov,Iwasaki,vDVmass2}. The Pauli-Fierz pure spin 2 theory has 5 degrees of freedom per spatial point (at least to lowest order), avoiding a wrong-sign spin 0 field by making it infinitely massive.  But the van Dam-Veltman-Zakharov discontinuity shows (at least perturbatively) an observable discontinuity in the massless limit, by which pure spin 2 massive gravities already were experimentally excluded: light spin 2 graviton theories do not approximate Einstein's equations.  The other possibility, involving spin 2 and a wrong-sign spin 0 at every point, is not expected to have such a discontinuity, but instead is believed to be violently unstable.  	The desire to avoid negative-energy degrees of freedom  motivates gauge invariance for linear theories \cite{DeserFermion,VanN} and thus leads to massless spin 2, that is, Einstein's equations.

 In the last decade or so, however, the consensus has weakened in both cases, partly because dark energy has provided empirical evidence that Einstein's equations might be wrong on long distance scales and partly because explanations of dark energy tend to involve weakening assumptions of energy positivity.  One finds claims and counterclaims regarding the non-perturbative smoothness of the massless limit of Pauli-Fierz pure spin 2 theories  \cite{Vainshtein,Vainshtein2,VainshteinReview,DeffayetStrong,DeffayetRecoveringPRL,DeffayetRecoveryPRD,Arkani,deRhamGabadadze,Gruzinov,Hinterbichler}.  The infinite-mass cases below are of this sort.  
One also finds both  increasingly frequent favorable mentions of spin 2-spin 0 theories   \cite{Visser,HamamotoMassless1,ScharfSecond,GrishchukMass,PetrovMass,SolovievMassive} and, sometimes independently,   increasing doubts that negative energy degrees of freedom are always bad.  Two considerations warrant giving spin 2-spin 0 theories enough consideration to assess by detailed calculation.  First, the old consensus against such theories appears to be motivated partly by a false converse of a true theorem.  While positive energy is sufficient for stability in mechanics, demonstrably it is \emph{not} necessary, as is known  in nonlinear stability theory and plasma and fluid physics  \cite{WeilandWilhelmsson,KhazinShnolBook,MeyerSchmidtArnold,MeyerHallOffin,CabralStability}.  Much depends on the resonances and nonlinear interaction terms, at least for classical finite-dimensional theories and field theories in a box (which seem to be the state of the mathematical art), and presumably on the dispersion relation(s) as well. Of course things might well be harder in field theory (in infinite space), and harder still under quantization.  It is noteworthy that  one also sees Kolmogorov-Arnol'd-Moser-type theorems being proved for partial differential equations  \cite{KuksinHamiltonianPDEs,GengKAMhigherdimPDE}.   But clearly negative energy degrees of freedom do at least make the question of stability significant. The second consideration is that there are (previously unnoticed, evidently) reasons to wonder whether there really are negative energy degrees of freedom at high wavenumbers.  While the claim that the spin 0 has negative energy goes back to Pauli and Fierz, the calculation turns out to be  nontrivial.  For the Hamiltonian in terms of the true degrees of freedom, taking the simplest cases with quadratic dependence on the shift vector \cite{MassiveGravity1,DeserMass,Marzban}, eliminating the shift vector using its equation of motion gives a positive cross-term $ \sim \mathcal{H}_i^2$ mixing the wrong-sign spin 0 and right-sign helicity 0 part of the spin 2.  The reduced Hamiltonian density contains not only derivatives of the canonical momenta due to eliminating the shift, but also higher spatial derivatives due to eliminating the lapse, and so does not fit the usual form. (The Lagrangian density, by contrast, becomes formally nonlocal, much as in (\cite{Marzban}), and hence not clearly simpler.)  Apart from detailed calculation, it isn't just obvious that there really are negative energy degrees of freedom  at high wavenumbers, where the strange new terms will be important.  Negative energies for low wavelengths only, it has been suggested for other theories, might permit  metastability   \cite{RubakovPhantom,SmilgaGhost,IzumiMassiveGhost}.  There are reasons to worry about negative energy problems, but not such compelling reasons as to obviate detailed investigation.  The theories derived here include both pure spin 2  and spin 2-spin 0 examples, so resolving the objections to either sort of theory would vindicate some of these theories.


\section{Curvature from Universal Coupling and Gauge Invariance: Covector Density Potential}

One can consider what theories result from a covector potential (with an additional local Lorentz index) with free field gauge freedom and coupling to the total stress energy tensor.  Any density weight will be permitted, except for a singular case.  
The massless case will be treated first, yielding Einstein's equations.  Later a mass term will be introduced. 


\subsection{Free Field Action }

	For the massless theories, one assumes an initial infinitesimal invariance (up to a boundary term) of the free gravitational action, along with local Lorentz freedom.  For the later derivation of massive theories, the gauge freedom will be broken by a term algebraic in the fields, but the derivative terms will retain the gauge invariance.  The entire theory will retain local Lorentz freedom.  

     Let $S_{f}$ be the action for a free covector  tensor density $\tilde{\gamma}_{\mu}^A$ (of density weight $-w,$   
$w \neq \frac{1}{4}$) in a spacetime with a flat metric tensor $\eta_{\mu\nu}$ in arbitrary coordinates in four space-time dimensions. 
The flat metric tensor
$\eta_{\mu\nu}$ is equivalent as a geometric object  to the weight $-2w$ tensor density $\tilde{\eta}_{\mu\nu}$; the two are related by \begin{equation} 
\tilde{\eta}_{\mu\nu} = \eta_{\mu\nu}  \sqrt{-\eta} \: ^{-2w}.
\end{equation}
 One notes that the forbidden case    $w \neq \frac{1}{4}$ makes   $\tilde{\eta}_{\mu\nu} $ non-invertible as a function of the metric: $\eta_{\mu\nu}  \sqrt{-\eta} \: ^{-1/2}$ determines only the null cone, not a full metric tensor. 
The weight $-2w$ metric density $\tilde{\eta}_{\mu\nu}$ can be built from the weight $-w$ cotetrad density $\tilde{\eta}_{\mu}^A$ by the relation
\begin{equation} 
\tilde{\eta}_{\mu\nu} = \tilde{\eta}_{\mu}^A   \eta_{AB}     \tilde{\eta}_{\nu}^B,
\end{equation}
where $\eta_{AB}$ is just the matrix $diag(-1,1,1,1).$  
$\tilde{\eta}_{\mu}^A $ will represent the geometry of space-time in this derivation.  One can therefore write the action for the free gravitational field as $S_f[\tilde{\gamma}_{\mu}^A, \tilde{\eta}_{\mu}^A].$   This action also involves the constant matrix $\eta_{AB}.$

The torsion-free metric-compatible covariant derivative is denoted by $\partial_{\mu}$, so $\partial_{\alpha} \eta_{\mu\nu} = 0$ and $\partial_{\alpha} \tilde{\eta}_{\mu\nu} = 0$.
One recalls that covariant and Lie differentiation of densities involves an extra term due to density weight.  A $(1, 1)$ density $\tilde{ \phi }^{\alpha}_{\beta}$ of weight $v$ is representative.  The Lie derivative is given by \cite{Schouten}
\begin{eqnarray}
\pounds_{\xi} \tilde{ \phi }^{\alpha}_{\beta} = \xi^{\mu} \tilde{ \phi } ^{\alpha}_{\beta},_{\mu} -
\tilde{ \phi }^{\mu}_{\beta} \xi^{\alpha},_{\mu} + \tilde{ \phi }^{\alpha}_{\mu} \xi^{\mu},_{\beta} + v
\tilde{ \phi }^{\alpha}_{\beta} \xi^{\mu},_{\mu}.
\end{eqnarray} 
The $\eta$-covariant derivative is given by  \cite{Schouten} 
\begin{eqnarray}   
\partial_{\mu} \tilde{ \phi }^{\alpha}_{\beta} = \tilde{ \phi }^{\alpha}_{\beta},_{\mu} + \tilde{ \phi }^{\sigma}_{\beta}
\Gamma_{\sigma\mu}^{\alpha} - \tilde{ \phi } ^{\alpha}_{\sigma} \Gamma_{\beta\mu}^{\sigma} - v
\tilde{ \phi }^{\alpha}_{\beta} \Gamma_{\sigma\mu}^{\sigma}.
\end{eqnarray} 
Here $\Gamma_{\beta\mu}^{\sigma}$ are the Christoffel symbols for $\eta_{\mu\nu}.$  Once the curved metric $g_{\mu\nu}$ is defined below, the analogous $g$-covariant derivative $\nabla$ with Christoffel symbols $\{ _{\sigma\mu}^{\alpha} \}$ follows.   
On account of the local Lorentz freedom, it is not the case that $\partial_{\alpha} \tilde{\eta}_{\mu}^A $ vanishes. While one could define another covariant derivative that corrects this problem, such a construction will not be needed here. 
 For the densitized flat metric  $\tilde{\eta}_{\mu\nu},$ the oppositely densitized inverse flat metric is  $\tilde{\eta}^{\mu\nu}.$  Likewise, the inverse of the densitized curved metric  $\tilde{g}_{\mu\nu}$  to be defined below is  $\tilde{g}^{\mu\nu}.$

 The potential problem of wrong-sign degrees of freedom can be avoided for massless theories using suitable gauge freedom.  Let us require that the free field action $S_{f}$ change only by a boundary term under the infinitesimal gauge transformation 
$ \tilde{\gamma}_{\mu}^A \rightarrow \tilde{\gamma}_{\mu}^A  + \delta \tilde{\gamma}_{\mu}^A$, where
\begin{eqnarray}
\delta \tilde{\gamma}_{\mu}^A = \tilde{\eta}_{\nu}^A \partial_{\mu} \xi^{\nu} +   
 c \tilde{\eta}_{\mu}^A \partial_{\nu} {\xi}^{\nu}.
\label{gaugeinv}
\end{eqnarray}
  One anticipates that a connection between $w$ and $c$ will emerge.  
This expression does not include the possible term $\xi^{\nu} \partial_{\nu} \tilde{\eta}_{\mu}^A,$ which would spoil local Lorentz invariance of this gauge transformation and also the form of the resulting generalized Bianchi identity as the divergence of an expression linear in $\frac{ \delta S_f}{\delta \tilde{\eta}_{\mu}^A }.$  Neither does it include 
$\tilde{\eta}_{\mu\rho} \tilde{\eta}^{\nu}_{B} \eta^{AB} \partial_{\nu} \xi^{\rho},$ which does not appear to be helpful for arriving eventually at an expression of the form $\delta \tilde{g}_{\mu\nu} \sim  \pounds_{\xi} \tilde{g}_{\mu\nu}.$

	    For any $S_{f}$ (quasi-)invariant in this sense under (\ref{gaugeinv}), a certain linear combination  the free field equations is
identically divergenceless.  The action changes by 
\begin{eqnarray}
\delta S_{f} = \int d^{4}x  \left [\frac{\delta S_{f} }{ \delta \tilde{\gamma}_{\mu}^A }  (\tilde{\eta}_{\nu}^A \partial_{\mu} \xi^{\nu} +   c \tilde{\eta}_{\mu}^A \partial_{\nu} \xi^{\nu}) \right]
\end{eqnarray} plus an irrelevant boundary term.   
 Integrating by parts,
letting $\xi^{\mu}$ have compact support to annihilate the boundary terms (as we shall do throughout the paper), and
making use of the arbitrariness of $\xi^{\mu}$, one obtains the identity
\begin{eqnarray}
- \partial_{\mu} \left(  \tilde{\eta}_{\nu}^A  \frac{\delta S_{f} }{  \delta \tilde{\gamma}_{\mu}^A }  + c   \delta^{\mu}_{\nu} \tilde{\eta}_{\alpha}^A   \frac{\delta S_{f} }{ \delta \tilde{\gamma}_{\alpha}^A }     \right) = 0.
\end{eqnarray}
This is the generalized Bianchi identity for the free theory on account of the gauge freedom.

Besides this identity, there is also the local Lorentz invariance of the free theory (not to mention the full theory) to consider.  The free field action $S_{f},$  under an infinitesimal local Lorentz transformation
\begin{eqnarray}
\delta \tilde{\gamma}_{\mu}^A = \Omega^{A}_{B} \tilde{\gamma}_{\mu}^B , \nonumber \\
\delta \tilde{\eta}_{\mu}^A = \Omega^{A}_{B} \tilde{\eta}_{\mu}^B, 
\end{eqnarray}
is left unchanged.  The matrix field $ \Omega^{A}_{B}$ is antisymmetric when an index is moved by $\eta_{AB}$ or its inverse.  
The resulting generalized Bianchi identity is
\begin{equation}
\tilde{\gamma}_{\mu [B} \frac{ \delta S_f }{ \delta \tilde{\gamma}_{\mu}^{A]} } +\tilde{\eta}_{\mu [B} \frac{ \delta S_f }{ \delta \tilde{\eta}_{\mu}^{A]} } =0.
\end{equation}



\subsection{Cotetrad Stress-Energy Tensor Density}

     If the energy-momentum tensor is to be the source for the gravitational potential $\tilde{\gamma}_{\mu}^A$,
consistency requires that the \emph{total} energy-momentum tensor be used, including
gravitational energy-momentum, not merely non-gravitational (``matter'') energy-momentum, for
only the total energy-momentum tensor is divergenceless in the sense of $\partial_{\nu}$
\cite{Deser,DeserVierbein}, or, equivalently, in the sense of a Cartesian coordinate divergence.  To obtain a
global conservation law, one needs a vanishing \emph{coordinate} divergence for the 4-current. 
Thus a vanishing covariant divergence (in terms of a curved connection) for a two-index energy-momentum complex is not useful.

	An expression for the total energy-momentum tensor density can be derived from $S$ using a tetrad (or cotetrad, or densitized tetrad, or densitized cotetrad) that corresponds to a flat metric tensor $\eta_{\mu\nu};$ for brevity, the entity will be called a flat (co)tetrad, even if it is densitized.  
One has $\tilde{\eta}_{\mu\nu} = \tilde{\eta}_{\mu}^A \eta_{AB} \tilde{\eta}_{\nu}^B,$ where $\eta_{AB}=diag(-1,1,1,1).$ 
Let the action $S$ depend on the flat cotetrad $\tilde{\eta}_{\mu}^A$, the (asymmetric) gravitational potential $\tilde{\gamma}_{\mu}^A$, and matter fields $u$.  Here $u$ represents an arbitrary collection of dynamical geometric objects (in the sense of having a set of components at each point in each local chart and a transformation rule in the overlap regions \cite{Anderson,Trautman}), having perhaps some coordinate indices but no Lorentz indices.    Under an arbitrary infinitesimal change of coordinates with compact support (so that boundary terms can be discarded without notice), described by a vector field
$\xi^{\mu}$, the action changes by the amount 
\begin{eqnarray}
\delta S = \int d^{4}x \left( \frac{\delta S}{\delta
\tilde{\gamma}_{\mu}^A} \pounds_{\xi} \tilde{\gamma}_{\mu}^A  +
\frac{\delta S}{\delta u} \pounds_{\xi} u +
\frac{\delta S}{\delta \tilde{\eta}_{\mu}^A } \pounds_{\xi} \tilde{\eta}_{\mu}^A \right) + BT,
\end{eqnarray}  
with boundary term $BT$  vanishing due to compact support of $\xi^{\mu}.$
But $S$ is a scalar, so $\delta S= 0$. 
Integrating by parts, discarding more vanishing boundary terms, and using the arbitrariness of the vector field $\xi^{\mu}$  gives a generalized Bianchi identity, from which the conservation of stress-energy can be inferred.  To have a moderately nontrivial sort of matter for definiteness, let $u$ be a single contravariant vector field $u^{\mu}.$
The resulting identity is 
\begin{eqnarray}
 \frac{\delta S}{\delta \tilde{\eta}_{\mu}^A } \partial_{\alpha} \tilde{\eta}_{\mu}^A   
-  \partial_{\mu} \left(  \tilde{\eta}_{\alpha}^A  \frac{\delta S}{\delta \tilde{\eta}_{\mu}^A }     
-  w \delta^{\mu}_{\alpha} \tilde{\eta}_{\rho}^A  \frac{\delta S}{\delta \tilde{\eta}_{\rho}^A }    \right)  \\ \nonumber
+ \frac{\delta S}{\delta \tilde{\gamma}_{\mu}^A } \partial_{\alpha} \tilde{\gamma}_{\mu}^A   
-  \partial_{\mu} \left(  \tilde{\gamma}_{\alpha}^A  \frac{\delta S}{\delta \tilde{\gamma}_{\mu}^A }     
-  w \delta^{\mu}_{\alpha} \tilde{\gamma}_{\rho}^A  \frac{\delta S}{\delta \tilde{\gamma}_{\rho}^A }    \right) 
+\frac{\delta S}{\delta u^{\mu} } \partial_{\alpha} u^{\mu}   + \partial_{\mu} \left( \frac{\delta S}{\delta u^{\alpha} } u^{\mu}    \right) 
  = 0.
\end{eqnarray}
The first term $\frac{\delta S}{\delta \tilde{\eta}_{\mu}^A } \partial_{\alpha} \tilde{\eta}_{\mu}^A    $
is neither part of a divergence nor vanishing on-shell (that is, using the matter and gravitational equations of motion); it also violates local Lorentz invariance.  Thus  it will need further attention.

For the local Lorentz freedom, a similar derivation to that for $S_f$ above shows that
\begin{equation}
\tilde{\gamma}_{\mu [B} \frac{ \delta S }{ \delta \tilde{\gamma}_{\mu}^{A]} } +\tilde{\eta}_{\mu [B} \frac{ \delta S }{ \delta \tilde{\eta}_{\mu}^{A]} } =0,  \end{equation} 
which implies that the antisymmetric part of the stress-energy tensor vanishes when gravity is on-shell.  Using the local Lorentz identity in the coordinate identity and rearranging some terms gives 
\begin{eqnarray}
\tilde{\gamma}_{\alpha[B} \frac{\delta S}{\delta \tilde{\gamma}_{\alpha}^{A]} } \tilde{\eta}^{B\mu} \partial_{\alpha} \tilde{\eta}_{\mu}^A   
-  \partial_{\mu} \left(  \tilde{\eta}_{\alpha}^A  \frac{\delta S}{\delta \tilde{\eta}_{\mu}^A }     
-  w \delta^{\mu}_{\alpha} \tilde{\eta}_{\rho}^A  \frac{\delta S}{\delta \tilde{\eta}_{\rho}^A }    \right)  \\ \nonumber
+ \frac{\delta S}{\delta \tilde{\gamma}_{\mu}^A } \partial_{\alpha} \tilde{\gamma}_{\mu}^A   
-  \partial_{\mu} \left(  \tilde{\gamma}_{\alpha}^A  \frac{\delta S}{\delta \tilde{\gamma}_{\mu}^A }     
-  w \delta^{\mu}_{\alpha} \tilde{\gamma}_{\rho}^A  \frac{\delta S}{\delta \tilde{\gamma}_{\rho}^A }    \right) 
+\frac{\delta S}{\delta u^{\mu} } \partial_{\alpha} u^{\mu}   + \partial_{\mu} \left( \frac{\delta S}{\delta u^{\alpha} } u^{\mu}    \right) 
  = 0.
\end{eqnarray} 
Letting matter $u$ and gravity $\tilde{\gamma}_{\mu}^A $ satisfy their equations of motion gives a conserved symmetric stress-energy tensor 
\begin{eqnarray}
-  \partial_{\mu} \left(  \tilde{\eta}_{\alpha}^A  \frac{\delta S}{\delta \tilde{\eta}_{\mu}^A }     
-  w \delta^{\mu}_{\alpha} \tilde{\eta}_{\rho}^A  \frac{\delta S}{\delta \tilde{\eta}_{\rho}^A }    \right) 
  = 0.
\end{eqnarray} 
This quantity is an energy-momentum tensor density for matter and gravitational fields. As usual with Rosenfeld-style derivations using a derivative with respect to a flat metric, one  relaxes  flatness  while taking the functional derivative and then restores flatness later \cite{RosenfeldStress,GotayMarsden}.


\subsection{Full Universally-Coupled Action}
	We seek an action $S$ obeying the plausible physical postulate that (invertible  linear combinations of) the Euler-Lagrange equations be just (invertible linear combinations of) the free field equations for $S_{f}$ augmented by the total energy-momentum tensor.
A simple way to impose this requirement using a densitized covector to represent the flat space-time metric and another one to represent the gravitational potential is:
\begin{eqnarray}
\frac{\delta S}{\delta \tilde{\gamma}_{\mu}^A } = \frac{\delta S_{f} }{\delta \tilde{\gamma}_{\mu}^A } -
\frac{\lambda }{2} \frac{\delta S}{\delta \tilde{\eta}_{\mu}^A }, 
\end{eqnarray}  
where $\lambda = - \sqrt{32 \pi G}$. 
The expression on the right involves a trace-altered relative of the stress-energy tensor.  (One could shift the trace and obtain the stress-energy tensor on the right side and a trace-altered relative of the Euler-Lagrange equations if one wished.)
One is free to make a change of variables in
$S$ from $\tilde{\gamma}_{\mu}^A$ and $\tilde{\eta}_{\mu}^A$  to the `bimetric' (really ``bitetrad,'' but that word is not catchy) variables
$\tilde{g}_{\mu}^A$ and $\tilde{\eta}_{\mu}^A$, where
\begin{eqnarray}
\tilde{g}_{\mu}^A = \tilde{\eta}_{\mu}^A  - \frac{\lambda}{2} \tilde{\gamma}_{\mu}^A.
\end{eqnarray} 
(From $\tilde{g}_{\mu}^A$ one can then define the metric $g_{\mu\nu}$ by matrix algebra and then define the $g$-covariant derivative $\nabla$ in the usual way.  We shall have little need for explicit use of   $\nabla$, however.)
 Equating coefficients of the variations using the old variables and the new variables in terms of the old gives
\begin{eqnarray}
 \frac{\delta S}{\delta \tilde{\eta}_{\mu}^A}|\tilde{\gamma} =   \frac{\delta S}{\delta
\tilde{\eta}_{\mu}^A} |\tilde{g}  +  \frac{\delta S}{\delta \tilde{g}_{\mu}^A}  
\end{eqnarray}
for $\delta \tilde{\eta}_{\mu}^A$
and 
\begin{eqnarray}   
 \frac{\delta S}{\delta \tilde{\gamma}_{\mu}^A}  =  - \frac{\lambda}{2} \frac{\delta S}{\delta \tilde{g}_{\mu}^A} 
\end{eqnarray}
for $\delta \tilde{\gamma}_{\mu}^A.$
Putting these two results together gives
\begin{eqnarray}
 \lambda \frac{\delta S}{\delta \tilde{\eta}_{\mu}^A}|\tilde{\gamma} =  \lambda   \frac{\delta S}{\delta \tilde{\eta}_{\mu}^A}|\tilde{g}  - {2} \frac{\delta S}{\delta \tilde{\gamma}_{\mu}^A}_,  
\end{eqnarray} 
which  splits the stress-energy tensor into one piece that vanishes when gravity is
on-shell and one piece that does not.  Using this result in the universal coupling postulate 
gives 
\begin{eqnarray}
\frac{\lambda}{2} \frac{\delta S}{\delta \tilde{\eta}_{\mu}^A}|\tilde{g} = \frac{\delta S_{f}}{\delta \tilde{\gamma} _{\mu}^A}.
 \end{eqnarray}

Multiplying by the flat tetrad, taking the divergence, and recalling the free field theory's Bianchi identity from gauge invariance gives
\begin{eqnarray}
- \frac{\lambda}{2} \partial_{\mu} \left(  \tilde{\eta}_{\nu}^A  \frac{\delta S }{  \delta \tilde{\eta}_{\mu}^A }|\tilde{g}  + c   \delta^{\mu}_{\nu} \tilde{\eta}_{\alpha}^A   \frac{\delta S }{ \delta \tilde{\eta}_{\alpha}^A }|\tilde{g}     \right) = 0.
\end{eqnarray}
Above $c$ was a free parameter, so it now makes sense, in view of the form of the stress-energy tensor, to set  $c=-w$.  By playing with the density weights to account for the trace term, one can show that the resulting equation is, in terms of the non-weighted tetrad $\eta_{\mu}^A$ (with no $\tilde{}$), 
\begin{eqnarray}
 \partial_{\mu} \left( \eta_{\nu}^A  \frac{\delta S }{  \delta \eta_{\mu}^A }|\tilde{g}  \right) = 0.
\end{eqnarray}
Raising an index gives 
\begin{eqnarray}
 \partial_{\mu} \left( \eta^{\nu A}  \frac{\delta S }{  \delta \eta_{\mu}^A }|\tilde{g}  \right) = 0.
\end{eqnarray}
There being no spinors present in the theory, one can choose for the curved tetrad $\tilde{g}_{\mu}^A$ to appear only through the curved metric $g_{\mu\nu},$ as one knows occurs in Einstein's equations. Local Lorentz invariance does not exclude dependence on, for example, $\tilde{g}^A_{\mu} \tilde{\eta}_A^{\nu}.$  Such dependence will appear below in the mass term as well.  But it is difficult to envision how a term that depends on both the curved and flat tetrads could satisfy the previous equation as an \emph{identity}. 
So restricting the action to depend on the curved tetrad only through the curved metric  might not be an additional assumption after all.  With dependence on the curved tetrad restricted to dependence through the curved metric, it follows as a theorem that dependence on the flat tetrad is also only through dependence on the corresponding flat metric $\eta_{\mu\nu}$ \cite{McKellarConcomitant}.  
One then has 
\begin{eqnarray}
 \partial_{\mu} \left( \eta^{\nu A}  \frac{\delta S }{  \delta \eta_{\mu}^A }|\tilde{g}  \right) = 2 \partial_{\mu}\frac{ \delta S}{\eta_{\mu\nu}}|\tilde{g} =0.
\end{eqnarray}
 One might wonder what has been gained using the tetrad formalism instead of a metric formalism.  (A tempting wrong answer is that one acquires the ability to have spinors in the theory. But spinors were already admissible at the cost of technical complexity and unfamiliarity, as noted above.)    The benefit will appear when the mass term is introduced, because the tetrad or cotetrad will permit a simple derivation of a mass term.  The density weight of the tetrad and the type (contravariant or covariant) of tetrad will determine the specific mass term obtained.

 The flat metric appears, after some changes of variables, only in a (symmetric)  expression with identically vanishing divergence.  This fact, combined with the split of the stress-energy tensor into two parts, shows that the gravitational field equations \emph{alone} entail conservation of energy-momentum,
without any separate postulation of the matter equations.
    The quantity $\frac{\delta S}{ \delta \eta _{\mu\nu}} |\tilde{g}$, being symmetrical and having identically vanishing divergence on either index, necessarily has the form 
\begin{eqnarray}   
\frac{\delta S}{ \delta \eta _{\mu\nu}} |\tilde{g} = \frac{1}{2}  \partial_{\rho} \partial_{\sigma} \left( 
{\mathcal{M}}
^{[\mu\rho][\sigma\nu]} +   {\mathcal{M}}
^{[\nu\rho][\sigma\mu]} \right)  + B \sqrt{-\eta} \eta^{\mu\nu}
\end{eqnarray}
\cite{Wald} (pp. 89, 429), where ${\mathcal{M}} ^{\mu\rho\sigma\nu}$ is a tensor
density of weight $1$ and $B$ is a constant.  This result follows from the converse of
Poincar\'{e}'s lemma in Minkowski spacetime. ${\mathcal{M}} ^{\mu\rho\sigma\nu}$ cannot be chosen arbitrarily, but rather must be chosen so that the term $\frac{\delta S_{f}}{\delta \tilde{\gamma} _{\mu\nu}}$ is accommodated.

 Gathering all dependence on
$\eta_{\mu\nu}$ (with $\tilde{g}_{\mu\nu},$ or equivalently $g_{\mu\nu},$ independent) into one term yields 
$S = S_{1} [\tilde{g}_{\mu\nu}, u] + S_{2}[\tilde{g}_{\mu\nu}, \eta_{\mu\nu}, u].
$   Using the effective curved metric density $\tilde{g}_{\mu\nu},$ one can define an effective curved metric  by 
$\tilde{g}_{\mu\nu} = g_{\mu\nu} \sqrt{-g} \: ^{-l}$ (where $l =2w$) and an inverse curved metric density 
$ \tilde{g}^{\mu\nu}.$
For $S_{1}$, we choose the Hilbert action for general relativity plus
minimally coupled matter and a cosmological constant:
\begin{eqnarray}   
S_{1} = \frac{1}{16 \pi G} \int d^{4}x \sqrt{-g} R(g)  - \frac{\Lambda}{8 \pi G} \int d^{4}x
\sqrt{-g} + S_{mt}[g_{\mu\nu}, u].
\end{eqnarray}
As is well-known, the Hilbert action is the simplest (scalar) action that can be constructed using
only the metric tensor. (One could also admit terms with higher derivatives of the curved metric.) If the gravitational field vanishes everywhere, then the gravitational action
ought to vanish also.  In the massless case, the result is that  $B=\Lambda/16 \pi G$.  For the massive generalization below, the gauge-breaking part of the mass term will introduce another zeroth order contribution that also needs canceling. One could also let the matter couple to the Riemann tensor for $g_{\mu\nu}$ or allow higher powers of the Riemann tensor into the gravitational action, if one wished.  For the massless case, one might  set $\Lambda=0$ \cite{FMS}. 
One easily verifies that if 
\begin{eqnarray}   
S_{2} = \frac{1}{2} \int d^{4}x R_{\mu\nu\rho\sigma} (\eta)
{\mathcal{M}} ^{\mu\nu\rho\sigma} (\eta_{\mu\nu}, g_{\mu\nu}, u ) \nonumber \\ + \int
d^{4}x \alpha^{\mu},_{\mu} + 2 B \int d^{4}x \sqrt{-\eta}, 
\end{eqnarray}
then $ \frac{\delta S_{2} }{  \delta \eta _{\mu\nu}}  |g $ has just the desired form, while $S_{2}$ does not affect the Euler-Lagrange equations because $ \frac{\delta S_{2} }{  \delta \tilde{g} _{\mu\nu}} =0$ and $ \frac{\delta S_{2} }{  \delta u } =0$ identically \cite{Kraichnan}. 
The coefficient $B$ of the 4-volume term is naturally chosen to cancel any other zeroth order term (such as from a cosmological constant) in the action, so
that the action vanishes when there is no gravitational field. The boundary term is at our disposal; if  $\alpha^{\mu}$ is a weight $1$ vector density, then  $S$ is  a coordinate scalar. 
  It is doubtful that the unmodified Hilbert action  is best, given its badly behaved conservation laws with the factor of $2$ problem \cite{KatzBicakLB,PetrovKatz2}. 
In summary,  the universally coupled action for the massless cotetrad density  is  
\begin{eqnarray}   
S = \frac{1}{16 \pi G} \int d^{4}x \sqrt{-g} R(g)  - \frac{\Lambda}{8 \pi G} \int d^{4}x
\sqrt{-g} + S_{mt}[g_{\mu\nu}, u] \nonumber \\
 + \frac{1}{2} \int d^{4}x R_{\mu\nu\rho\sigma} (\eta)
{\mathcal{M}} ^{\mu\nu\rho\sigma} + \frac{\Lambda}{8 \pi G}  \int d^{4}x \sqrt{-\eta} \nonumber \\ + \int d^{4}x  \alpha^{\mu},_{\mu},
\end{eqnarray}
for which the first few terms give Einstein's equations and the last few give no contribution to the field equations. 



\section{Massive  Cotetrad  Density Theories}

	The present goal is to generalize the derivation above to yield one or more massive finite-range variants of Einstein's equations.  Such field equations would relate to Einstein's in much the way that Proca's massive electromagnetic field equations relate to Maxwell's \cite{Jackson}.  There are some differences, however, including the lack of uniqueness for massive gravity, even given the plausible postulate of universal coupling. It will turn out that every real number (with one exceptional forbidden case) gives a distinct massive variant of General Relativity.

One expects that the mass term for a free field be quadratic in the potential and lack derivatives.  The free field action $S_{f}$ is now assumed to have two parts: a (mostly kinetic) part $S_{f0}$ that is invariant under the erstwhile gauge transformations as in the massless case above, and an algebraic mass term $S_{fm}$ that is quadratic and breaks the gauge symmetry.  We seek a full universally coupled theory with an action $S$ that has two corresponding parts.  The two parts of $S=S_{0} + S_{ms}$ are the familiar part $S_{0}$ (yielding the Einstein tensor, the matter action, a cosmological constant, and  a zeroth order 4-volume term) and the new gauge-breaking part $S_{ms}$ which also has another zeroth order 4-volume term.  As it turns out, the mass term is built out of \emph{both} of the algebraic part of $S_{0} $ (the cosmological constant and 4-volume term) and the purely algebraic term $S_{ms}.$ Moreover, every part of the action is locally Lorentz-invariant.  

	Requiring $S_{f0}$ to change only by a boundary term under the variation  
$\delta \tilde{\gamma}_{\mu}^A  = \tilde{\eta}_{\nu}^A \partial_{\mu} \xi^{\nu}  -w \tilde{\eta}_{\mu}^A \partial_{\nu} \xi^{\nu}$  implies the identity
\begin{eqnarray}
- \partial_{\mu} \left(  \tilde{\eta}_{\nu}^A  \frac{\delta S_{f0} }{  \delta \tilde{\gamma}_{\mu}^A }  -w   \delta^{\mu}_{\nu} \tilde{\eta}_{\alpha}^A   \frac{\delta S_{f0} }{ \delta \tilde{\gamma}_{\alpha}^A }     \right) = 0.
\end{eqnarray}
Again we postulate universal coupling in the form
\begin{eqnarray}
\frac{\delta S}{\delta \tilde{\gamma}_{\mu}^A } = \frac{\delta S_{f} }{\delta \tilde{\gamma}_{\mu}^A } -
\frac{\lambda }{2} \frac{\delta S}{\delta \tilde{\eta}_{\mu}^A }, 
\end{eqnarray}  
Changing to the bimetric variables  implies, as before, that
\begin{eqnarray}
\frac{\lambda}{2} \frac{\delta S}{\delta \tilde{\eta}_{\mu}^A}|\tilde{g} = \frac{\delta S_{f}}{\delta \tilde{\gamma} _{\mu}^A}.
 \end{eqnarray}

	Now we  introduce the relations $S_{f}= S_{f0} + S_{fm}$ and $S=S_{0} + S_{ms}$ to treat separately the pieces that existed in the massless case from the innovations of the massive case.  Thus 
\begin{eqnarray}
\frac{\lambda}{2} \frac{\delta S_{0}}{\delta \tilde{\eta}_{\mu}^A}|\tilde{g} + \frac{\lambda}{2} \frac{\delta S_{ms}}{\delta \tilde{\eta}_{\mu}^A}|\tilde{g}= \frac{\delta S_{f0}}{\delta \tilde{\gamma} _{\mu}^A} + \frac{\delta S_{fm}}{\delta \tilde{\gamma} _{\mu}^A}.
 \end{eqnarray}
Given the assumption that the new terms $S_{fm}$ and $S_{ms}$ correspond, this equation separates into the familiar part $$\frac{\delta S_{f0}}{\delta \tilde{\gamma}_{\mu}^A} =  \frac{ \lambda}{2} \frac{\delta S_{0} }{\delta \tilde{\eta}_{\mu}^A}|\tilde{g} $$ and the new part 
$$ \frac{\delta S_{fm}}{\delta \tilde{\gamma}_{\mu}^A}  =  \frac{ \lambda}{2} \frac{\delta S_{ms} }{\delta \tilde{\eta}_{\mu}^A} | \tilde{g}.$$  Using the gauge invariance result for the massless \emph{part} of the free field action, \emph{etc.} as above, one derives the form of $S_{0} $   to be 
\begin{eqnarray}   
S_{0} = S_{1} [\tilde{g}_{\mu\nu}, u] +   S_{2},
\end{eqnarray}
with $S_2$ as in the massless case.


Assuming the free field mass term to be quadratic in the gravitational potential, 
one can assume the form
$$ S_{fm} = \sqrt{-\eta} \tilde{\gamma}^A_{\mu} \tilde{\gamma}^B_{\nu} ( C \eta_{AB} \tilde{\eta}^{\mu\nu} + D \tilde{\eta}^{\mu}_B \tilde{\eta}^{\nu}_A + E \tilde{\eta}^{\mu}_A  \tilde{\eta}^{\nu}_B)_.$$
Its variational derivative is
$$ \frac{\delta S_{fm}}{\delta \tilde{\gamma}_{\mu}^A}  = 2 \sqrt{-\eta} \tilde{\gamma}_{\nu}^B ( C \eta_{AB}  \tilde{\eta}^{\mu\nu}  + D \tilde{\eta}^{\mu}_B  \tilde{\eta}^{\nu}_A + E \tilde{\eta}^{\nu}_B  \tilde{\eta}^{\mu}_A)_.$$ Changing to the bimetric variables gives
\begin{equation}
\frac{\delta S_{fm}}{\delta \tilde{\gamma}_{\mu}^A}= \frac{4 \sqrt{-\eta}}{ \lambda}([C+D]\tilde{\eta}^{\mu}_A - C \tilde{g}^B_{\nu} \eta_{AB} \tilde{\eta}^{\mu\nu} - D \tilde{g}^B_{\nu} \tilde{\eta}^{\mu}_B \tilde{\eta}^{\nu}_A + E  \delta^{\nu}_{\nu} \tilde{\eta}^{\mu}_A - E \tilde{g}^B_{\nu} \tilde{\eta}_B^{\nu} \tilde{\eta}^{\mu}_A);
\end{equation} 
the term with  $\delta^{\nu}_{\nu}$ is a reminder that the detailed coefficients depend on the space-time dimension, which has already been assumed to be $4.$
The new part of the action for the massive case satisfies 
$$ \frac{\delta S_{fm}}{\delta \tilde{\gamma}_{\mu}^A}  =   \frac{ \lambda}{2} \frac{\delta S_{ms} }{\delta \tilde{\eta}_{\mu}^A} | \tilde{g},$$  so that becomes, using the explicit quadratic form of $S_{fm},$
$$\frac{4 \sqrt{-\eta}}{ \lambda}([C+D+ 4E]\tilde{\eta}^{\mu}_A - C \tilde{g}^B_{\nu} \eta_{AB} \tilde{\eta}^{\mu\nu} - D \tilde{g}^B_{\nu} \tilde{\eta}^{\mu}_B \tilde{\eta}^{\nu}_A - E \tilde{g}^B_{\nu} \tilde{\eta}_B^{\nu} \tilde{\eta}^{\mu}_A) =  \frac{ \lambda}{2} \frac{\delta S_{ms} }{\delta \tilde{\eta}_{\mu}^A} | \tilde{g}.$$ Using $det(\tilde{\eta}^A_{\mu}) = \sqrt{-\eta} \; ^{1-4w},$ one finds the useful result $$ \frac{ \partial  \sqrt{-\eta} }{ \partial \tilde{\eta}^A_{\mu} }=\frac{1}{1-4w} \sqrt{-\eta} \tilde{\eta}^{\mu}_{A.} $$

One now needs to try to find an appropriate form for  $S_{ms}$. It appears impossible to find anything to contribute a term of the form $C \tilde{g}^B_{\nu} \eta_{AB} \tilde{\eta}^{\mu\nu};$ one can show that this apparent impossibility is genuine, because the term is not an exact differential, whereas the remaining terms, with suitable coefficients, are.  Thus $C=0.$  For the remaining terms, a natural form to try is 
$S_{ms} = \int d^{4}x (P \tilde{g}_{\nu}^B \tilde{\eta}^{\nu}_B + Q) \sqrt{-\eta}$
for unspecified real numbers $P$ and $Q$ to be determined shortly.  One notes that a pure $\sqrt{-g}$ piece that gives a cosmological constant plays no role here in this derivative, and can be  already included in $S_{0}.$

Equating $ \frac{\lambda}{2} \frac{\delta S_{ms} }{\delta \tilde{\eta}_{\mu}^A} | \tilde{g}$  with $ \frac{\delta S_{fm}}{\delta \tilde{\gamma} _{\mu}^A}  $ and equating coefficients determines several of the constants.  Equating the coefficients of the $\sqrt{-\eta} \tilde{\eta}^{\mu}_A$ terms gives $Q = \frac{8}{32 \pi G}(D+4E)(1-4w)$ (recalling that $\lambda^2 = 32 \pi G$).  Equating the coefficients of the $\sqrt{-\eta} \tilde{\eta}^{\mu}_B \tilde{\eta}^{\nu}_A \tilde{g}_{\nu}^B$ terms gives $P=\frac{8D}{32 \pi G}.$  Equating the coefficients of the 
$\sqrt{-\eta} \tilde{\eta}^{\mu}_A \tilde{\eta}^{\nu}_B \tilde{g}_{\nu}^B$ terms gives  $P=\frac{-8E(1-4w)}{32 \pi G}.$   Using all three results together gives 
\begin{eqnarray} 
E=\frac{D}{4w-1}, \nonumber \\
P=\frac{D}{4 \pi G} \nonumber \\
Q= \frac{ -D(4w+3)}{4 \pi G}. \end{eqnarray}
Thus one has 
$$\mathcal{L}_{ms} = -\frac{(4w+3)D}{4 \pi G} \sqrt{-\eta} + \frac{D}{4 \pi G} \sqrt{-\eta} \tilde{g}^B_{\nu} \tilde{\eta}_B^{\nu} $$
and $$\mathcal{L}_{fm}= \sqrt{-\eta} \tilde{\gamma}^A_{\mu} \tilde{\gamma}^B_{\nu} (0 \eta_{AB} \tilde{\eta}^{\mu\nu} +  D \tilde{\eta}^{\mu}_B \tilde{\eta}^{\nu}_A + \frac{D}{4w-1} \tilde{\eta}^{\nu}_B \tilde{\eta}^{\mu}_A).$$

 Combining the algebraic piece of $S_{0}$ with $S_{ms}$ gives 
\begin{eqnarray}
S_{alg}= - \frac{\Lambda}{8 \pi G} \int d^4 x \sqrt{-g} + 2B\int d^4 x \sqrt{ -\eta}  \nonumber \\
+  \int d^4 x\left(  \frac{-(4w-3)D\sqrt{ -\eta}}{4 \pi G} + \frac{D\sqrt{ -\eta}}{4 \pi G} \tilde{g}^B_{\nu} \tilde{\eta}^{\nu}_B \right)_.
\end{eqnarray}
When the gravitational potential vanishes, $S_{alg}$ ought to vanish as well.  Imposing this condition to zeroth order gives
$B= \frac{\Lambda  + 8Dw -14D}{16 \pi G}.$
Recalling that the goal is to find a massive generalization of Einstein's theory (with no effective cosmological constant), we require the first-order term in $\tilde{\gamma}_{\mu}^A$ to vanish as well.  It follows that $\Lambda = 2D(1-4w)$ and $B=\frac{3D}{4 \pi G}.$   Thus the sign of the \emph{formal} cosmological constant term depends on the density weight of the potential chosen initially.  One also expects the quadratic part of the algebraic piece of the action $S_{alg}$ to agree with the free field mass term $S_{fm}.$  After a binomial expansion and some algebra, one sees that this is the case.
One has $$\mathcal{L}_{alg} = D \sqrt{-\eta} \left( \tilde{\gamma}^A_{\mu} \tilde{\eta}^{\mu}_B \tilde{\gamma}^B_{\nu} \tilde{\eta}^{\nu}_A  - \frac{ (\tilde{\gamma}^A_{\mu} \tilde{\eta}^{\mu}_A )^2 }{1-4w}   \right) + HOT,$$
where $HOT$ involves cubic and higher order terms.

   Making a weak-field expansion of the full massive nonlinear action $S$ to relate the coefficient $D$ to the mass $m$ of the spin 2 gravitons shows that $D=-\frac{m^2}{2}.$  For non-tachyonic theories, one imposes $D<0.$  A helpful fact in this expansion is that using the field redefinition 
$\tilde{g}_{\mu}^A = \tilde{\eta}_{\mu}^A  - \frac{\lambda}{2} \tilde{\gamma}_{\mu}^A,$ 
the perturbation of the associated metric is, to first order,
$$ \tilde{g}_{\mu\nu} = \tilde{\eta}_{\mu\nu}  - \lambda \tilde{\gamma}_{(\mu\nu)} + \ldots.$$ This choice of convention, inspired by (\cite{OP,OPMassive2}), makes the traceless part of the metric perturbation independent of the type of gravitational potential (\emph{e.g.}, covariant \emph{vs.} contravariant) as far as is readily possible.  This goal explains some of the factors of $\frac{1}{2}$ that appeared above.

	Combining all these results gives the total massive action $S,$ which depends on the spin 2 graviton mass $m$ and the density weight parameter $w$ which controls the relative mass of the spin 0 to that of the spin 2:
\begin{eqnarray}
S= \frac{1}{16 \pi G} \int d^{4}x \sqrt{-g} R(g)  + S_{matter}[\tilde{g}_{\mu\nu}, u]  \nonumber \\ + \frac{1}{2} \int d^{4}x R_{\mu\nu\rho\sigma} (\eta) 
 {\mathcal{M}}^{\mu\nu\rho\sigma}[\tilde{\eta}_{\mu\nu},\tilde{g}_{\mu\nu},u] + \int d^{4}x \partial_{\mu} \alpha^{\mu}       \nonumber   \\
+ \frac{m^2}{8 \pi G} \int d^{4}x ( \sqrt{-g}[1-4w] + \sqrt{-\eta}[4w+3] - \sqrt{- \eta} \, \tilde{g}_{\mu}^A \tilde{\eta}^{\mu}_A),  
\end{eqnarray}
for $w \neq \frac{1}{4}.$ Note that the vanishing of the coefficient $(1-4w)$ of the $\sqrt{-g}$ term would be a catastrophe and so is forbidden; its forbiddenness appeared at the beginning of the derivation, where that value would yield a densitized metric describing only the null cone, not the whole metric tensor, and so would involve coupling only to the traceless part of the stress-energy tensor \cite{SliBimGRG}.
By contrast,  the vanishing of the coefficient of the $\sqrt{-\eta}$ is permissible and implies conformal invariance.  The first line in the action contributes the ingredients of General Relativity, the second line contributes nothing to the field equations (but something to the Rosenfeld stress-energy tensor), and the third line contributes the mass term.
These theories are all universally coupled, \emph{pace} the claim \cite{FMS,DeserMass} 
that the Freund-Maheshwari-Schonberg theory has that feature uniquely.

One readily sees that, on fixing the local Lorentz freedom, these theories fall within the 2-parameter Ogievetsky-Polubarinov family \cite{OP}.  One needs to identify the parameter $w$ with their $-\frac{p}{2}$ and to notice that a cotetrad is roughly an inverse square root of the contravariant tensor $g^{\mu\nu}$ that they take as basic, so their parameter $n$ here takes the value of $-\frac{1}{2}.$  One can then use their calculation of the ratio of the spin $0$ mass $m_0$ to the spin $2$ mass $m$:
\begin{equation} m_0 = m \sqrt{ \frac{-16w^2 -8w + 3}{8w^2 + 4w} }_.
\end{equation}
(The calculation is mildly nontrivial because the density weight of the potential  affects the trace piece of the mass term.)  
Requiring the mass to be real (non-tachyonic) excludes a great many values, but leaves two intervals. (By contrast the densitized metric  theories  contribute a    single interval \cite{MassiveGravity1}.)  The mass ratio is shown in the figure. 
\begin{figure}
{\includegraphics[0in,0in][3.in,2in]{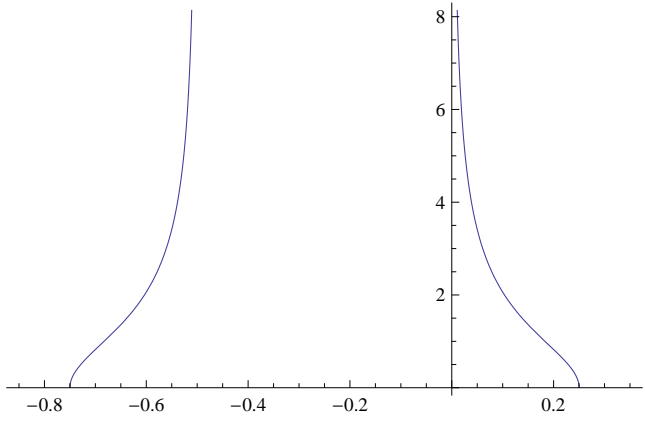}}
\caption{Ratio of Spin $0$ Mass to Spin $2$ Mass as Function of $w$; done with \emph{Mathematica}.}
\end{figure}
One recalls that the $\sqrt{-\eta}$ term may vanish, giving a massless spin $0$ and conformal invariance, but the $\sqrt{-g}$ term must not vanish.  The permitted intervals are $[-\frac{3}{4}, -\frac{1}{2}]$ and $[0, \frac{1}{4}).$  For $w= -\frac{3}{4},$ conformal invariance and masslessness for the spin $0$ obtain.  For $w= - \frac{1 + \sqrt{3} }{4} \approx -.683013,$ $m_0 = m.$ On this branch the spin $0$ mass rises with $w$; the value $w=-\frac{1}{2}$ gives an infinitely massive spin $0,$ as indicated by the vertical asymptote in the graph, and reduces to the Pauli-Fierz pure spin $2$ case to quadratic order.  The case $w=0$ on the other branch is similar. The cases $w=-\frac{1}{2}$ (weight $\frac{1}{2}$) and $w=0$ (weight $0$) correspond, respectively, to the fourth and second pure spin $2$ hadron mass terms of Zumino \cite[p. 492]{ZuminoDeser}.  
 From the vertical asymptote the mass descends, the ratio reaching $1$ at $w= \frac{-1 + \sqrt{3} }{2} \approx .183013$.  As $w$ approaches $\frac{1}{4}$ from the left, the spin $0$ mass goes to $0,$ but it never reaches $0$ because $ w \neq  \frac{1}{4}$.  Higher values of $w$ lead to imaginary masses.  In contrast to the two metric-based families previously derived by universal coupling \cite{MassiveGravity1}, the tetrad theories permit the spin $0$ mass to be heavier than the spin $2$.  Such flexibility expands the possible astrophysical and cosmological phenomenology  \cite{GrishchukMass,PetrovMass} of massive gravities
to theories derived using universal coupling.  This flexibility might also be relevant to the stability issue, which is expected to depend on resonances \cite{KhazinShnolBook} as appeared above; resonances in this case are determined by the ratio of the spin $0$ and spin $2$ masses. 


Reconsidering the generalized Bianchi identities for both coordinate and local Lorentz freedom in terms of the bimetric variables, one can find a shortcut to the inference of a Lorenz-Lorentz-type auxiliary condition, that is, one similar to the electromagnetic condition $\partial_{\mu} A^{\mu}=0.$  Using both identities and the gravitational \emph{and} matter field equations, one can infer that 
\begin{eqnarray}
\partial_{\mu} \left(  \tilde{\eta}_{\nu}^A  \frac{\delta S }{  \delta \tilde{\eta}_{\mu}^A }|\tilde{g}  -w  \delta^{\mu}_{\nu} \tilde{\eta}_{\alpha}^A   \frac{\delta S }{ \delta \tilde{\eta}_{\alpha}^A }|\tilde{g}     \right) = 0.
\end{eqnarray}
The massive theory's action splits into three parts, as noted above, which act differently here.  
The first part, $$ \frac{1}{16 \pi G} \int d^{4}x \sqrt{-g} R(g)  + S_{matter}[\tilde{g}_{\mu\nu}, u], $$ which contributes general relativistic terms to the field equations, contributes nothing here, because the flat tetrad is absent.  The second part
$$ \frac{1}{2} \int d^{4}x R_{\mu\nu\rho\sigma} (\eta)  {\mathcal{M}}^{\mu\nu\rho\sigma}[\tilde{\eta}_{\mu\nu},\tilde{g}_{\mu\nu},u] + \int d^{4}x \partial_{\mu} \alpha^{\mu},    $$
which contributes nothing to the field equations, contributes here a term with that is nonzero to 
 $\left(  \tilde{\eta}_{\nu}^A  \frac{\delta S }{  \delta \tilde{\eta}_{\mu}^A }|\tilde{g}  -w  \delta^{\mu}_{\nu} \tilde{\eta}_{\alpha}^A   \frac{\delta S }{ \delta \tilde{\eta}_{\alpha}^A }|\tilde{g}     \right),$ but its divergence vanishes identically.  
The third piece, the algebraic part, which gives the graviton masses, 
$$  \frac{m^2}{8 \pi G} \int d^{4}x ( \sqrt{-g}(1-4w) + \sqrt{-\eta}[4w+3] - \sqrt{- \eta} \, \tilde{g}_{\mu}^A \tilde{\eta}^{\mu}_A), $$ gives an interesting result:
\begin{equation}
\partial_{\mu} (\tilde{g}^{B}_{\nu} \tilde{\eta}^{\mu}_B  -[1+w] \tilde{g}^B_{\rho} \tilde{\eta}^{\rho}_B \delta^{\mu}_{\nu}) =0. \end{equation}  
This condition eliminates a spin $1$ and a spin $0$ from the field content.  It is closely analogous to the fact that the Lorenz-Lorentz condition $\partial_{\mu} A^{\mu}=0$ follows for massive Proca electromagnetism using the electromagnetic field equations and (to enforce charge conservation) the equations of motion for charged matter.  Ogievetsky and Polubarinov \cite{OP,OPMassive2}, who did not employ a tetrad formalism (but rather a surprisingly capacious formalism that takes powers of the metric using  binomial series expansions), derived massive gravities using such a spin limitation principle instead of universal coupling.

The mass term is the only place where the tetrads, as opposed to the metrics, appear essentially.  Thus the mass term is the only piece for which the local Lorentz invariance identity  gives interesting results.  That identity can be written as 
\begin{equation}
\tilde{g}_{\mu [B} \frac{ \delta S }{ \delta \tilde{g}_{\mu}^{A]} } +\tilde{\eta}_{\mu [B} \frac{ \delta S }{ \delta \tilde{\eta}_{\mu}^{A]} } =0.
\end{equation} 
Using the gravitational equations of motion gives
\begin{equation}
\tilde{\eta}_{\mu [B} \frac{ \delta S }{ \delta \tilde{\eta}_{\mu}^{A]} } =0.
\end{equation} 
Inserting the contribution from the mass term gives the result
\begin{equation}
\tilde{g}_{\rho[B} \tilde{\eta}_{A]}^{\rho} =0, 
\end{equation} 
or equivalently,
\begin{equation}
\tilde{g}_{[\mu A} \tilde{\eta}^{A}_{\nu]} =0; 
\end{equation}
the  antisymmetrization applies only over the Greek indices in this equation.  While the tetrads are asymmetric, expressions with either only Latin or only Greek indices are symmetric.


\section{Massless Universally Coupled Gravity with Densitized Tetrad }

I now turn to deriving first General Relativity, and then massive variants thereof, using the \emph{contravariant}  version of the derivation above.  There is not a complete symmetry on account of the fact that a covariant Lagrangian density has weight $1$, not weight $0$; thus one cannot simply swap indices up and down and interchange some $+$ and $-$ signs everywhere.  But conceptually the process is identical, so the exposition will be more concise.


\subsection{Massless Free Field Action }

     Let $S_{f}$ be the action for a set of four free covector   densities $\tilde{\gamma}_{\mu}^A$ of density weight $w,$   
$w \neq \frac{1}{4}$. 
The weight $2w$ densitized inverse  metric tensor
 $\tilde{\eta}^{\mu\nu}$ is essentially the densitized tetrad squared: 
\begin{equation}
\tilde{\eta}^{\mu\nu} = \tilde{\eta}^{\mu}_A   \eta^{AB}  \tilde{\eta}^{\nu}_B.
\end{equation}
It is the inverse of $\tilde{\eta}_{\mu\nu}$ for the same value of $w.$

  Let us require that the free field action $S_{f}$ change only by a boundary term under the infinitesimal gauge transformation 
\begin{eqnarray}
\delta \tilde{\gamma}^{\mu}_A = -\tilde{\eta}^{\nu}_A \partial_{\nu} \xi^{\mu} +   
 w \tilde{\eta}^{\mu}_A \partial_{\nu} {\xi}^{\nu}.  
\end{eqnarray}
  One obtains the identity
\begin{eqnarray}
 \partial_{\nu} \left(  \tilde{\eta}^{\nu}_A  \frac{\delta S_{f} }{  \delta \tilde{\gamma}^{\mu}_A }  -w   \delta_{\mu}^{\nu} \tilde{\eta}^{\alpha}_A   \frac{\delta S_{f} }{ \delta \tilde{\gamma}^{\alpha}_A }     \right) = 0.
\end{eqnarray}
There is also the local Lorentz invariance of the free theory:
\begin{equation}
\tilde{\gamma}^{\mu [B} \frac{ \delta S_f }{ \delta \tilde{\gamma}^{\mu}_{A]} } + \tilde{\eta}^{\mu [B} \frac{ \delta S_f }{ \delta \tilde{\eta}^{\mu}_{A]} } =0.
\end{equation}

Using  an arbitrary infinitesimal change of coordinates with compact support, one derives the generalized Bianchi identity for any invariant action.  Using the matter and gravitational equations of motion, as well as the other (local Lorentz) Bianchi identity, one arrives at an expression for a conserved symmetric stress-energy tensor (density):
\begin{eqnarray}
  \partial_{\nu} \left(  \tilde{\eta}^{\nu}_A  \frac{\delta S}{\delta \tilde{\eta}^{\mu}_A }     
-  w \delta^{\nu}_{\mu} \tilde{\eta}^{\alpha}_A  \frac{\delta S}{\delta \tilde{\eta}^{\alpha}_A }    \right) 
  = 0.
\end{eqnarray}

One can impose universal coupling with the postulate
\begin{eqnarray}
\frac{\delta S}{\delta \tilde{\gamma}^{\mu}_A } = \frac{\delta S_{f} }{\delta \tilde{\gamma}^{\mu}_A } +
\frac{\lambda }{2} \frac{\delta S}{\delta \tilde{\eta}^{\mu}_A. } 
\end{eqnarray}  
One is free to make a change of variables from $\tilde{\gamma}^{\mu}_A$ and $\tilde{\eta}^{\mu}_A$  to the `bimetric' variables
$\tilde{g}^{\mu}_A$ and $\tilde{\eta}^{\mu}_A$, where
\begin{eqnarray}
\tilde{g}^{\mu}_A = \tilde{\eta}^{\mu}_A  + \frac{\lambda}{2} \tilde{\gamma}^{\mu}_A.
\end{eqnarray} 
 Equating coefficients  gives
\begin{eqnarray}
 \frac{\delta S}{\delta \tilde{\eta}^{\mu}_A}|\tilde{\gamma} =   \frac{\delta S}{\delta
\tilde{\eta}^{\mu}_A} |\tilde{g}  +  \frac{\delta S}{\delta \tilde{g}^{\mu}_A}  
\end{eqnarray}
for $\delta \tilde{\eta}^{\mu}_A$
and 
\begin{eqnarray}   
 \frac{\delta S}{\delta \tilde{\gamma}^{\mu}_A}  =   \frac{\lambda}{2} \frac{\delta S}{\delta \tilde{g}^{\mu}_A} 
\end{eqnarray}
for $\delta \tilde{\gamma}^{\mu}_A.$
The universal coupling postulate becomes
\begin{eqnarray}
-\frac{\lambda}{2} \frac{\delta S}{\delta \tilde{\eta}^{\mu}_A}|\tilde{g} = \frac{\delta S_{f}}{\delta \tilde{\gamma}^{\mu}_A}.
 \end{eqnarray}

Multiplying by the flat tetrad, taking the divergence, and recalling the free field theory's Bianchi identity from gauge invariance gives
\begin{eqnarray}
 \partial_{\nu} \left(  \tilde{\eta}^{\nu}_A  \frac{\delta S }{ \delta \tilde{\eta}^{\mu}_A }|\tilde{g}  -w   \delta_{\mu}^{\nu} \tilde{\eta}^{\alpha}_A   \frac{\delta S }{ \delta \tilde{\eta}^{\alpha}_A }|\tilde{g}   \right) = 0.
\end{eqnarray}
One can show that the resulting equation is, in terms of the non-weighted tetrad $\eta^{\mu}_A$, 
\begin{eqnarray}
 \partial_{\nu} \left( \eta^{\nu}_A  \frac{\delta S }{ \delta \eta^{\mu}_A }|\tilde{g}  \right) = 0.
\end{eqnarray}
If one assumes that curved tetrad $\tilde{g}^{\mu}_A$ appears only through the curved metric $g_{\mu\nu},$ as one knows occurs in Einstein's equations, then the flat tetrad also appears only through its corresponding metric. By the same reasoning as above,
one obtains
\begin{eqnarray}   
S = \frac{1}{16 \pi G} \int d^{4}x \sqrt{-g} R(g)  - \frac{\Lambda}{8 \pi G} \int d^{4}x
\sqrt{-g} + S_{mt}[g_{\mu\nu}, u] \nonumber \\
 + \frac{1}{2} \int d^{4}x R_{\mu\nu\rho\sigma} (\eta)
{\mathcal{M}} ^{\mu\nu\rho\sigma} + \frac{\Lambda}{8 \pi G}  \int d^{4}x \sqrt{-\eta} \nonumber \\ + \int d^{4}x  \alpha^{\mu},_{\mu},
\end{eqnarray}
which gives Einstein's equations.


\section{Massive  Tetrad  Density Theories}

	The tetrad derivation can be generalized in the presence of a mass term.
Thus $S_f= S_{f0} + S_{fm}$ and $S=S_0 + S_{ms}.$ 
Requiring $S_{f0}$ to change only by a boundary term under the variation  
\begin{eqnarray}
\delta \tilde{\gamma}^{\mu}_A = -\tilde{\eta}^{\nu}_A \partial_{\nu} \xi^{\mu} +   
 w \tilde{\eta}^{\mu}_A \partial_{\nu} {\xi}^{\nu}.  
\end{eqnarray}
  One obtains the identity
\begin{eqnarray}
 \partial_{\nu} \left(  \tilde{\eta}^{\nu}_A  \frac{\delta S_{f0} }{  \delta \tilde{\gamma}^{\mu}_A }  -w   \delta_{\mu}^{\nu} \tilde{\eta}^{\alpha}_A   \frac{\delta S_{f0} }{ \delta \tilde{\gamma}^{\alpha}_A }     \right) = 0.
\end{eqnarray}
There is also the local Lorentz invariance of the free theory:
\begin{equation}
\tilde{\gamma}^{\mu [B} \frac{ \delta S_{f0} }{ \delta \tilde{\gamma}^{\mu}_{A]} } + \tilde{\eta}^{\mu [B} \frac{ \delta S_{f0} }{ \delta \tilde{\eta}^{\mu}_{A]} } =0.
\end{equation}

One can impose universal coupling with the postulate
\begin{eqnarray}
\frac{\delta S}{\delta \tilde{\gamma}^{\mu}_A } = \frac{\delta S_{f} }{\delta \tilde{\gamma}^{\mu}_A } +
\frac{\lambda }{2} \frac{\delta S}{\delta \tilde{\eta}^{\mu}_A. } 
\end{eqnarray}  
One is free to make a change of variables from $\tilde{\gamma}^{\mu}_A$ and $\tilde{\eta}^{\mu}_A$  to the `bimetric' variables
$\tilde{g}^{\mu}_A$ and $\tilde{\eta}^{\mu}_A$.
The universal coupling postulate becomes
\begin{eqnarray}
-\frac{\lambda}{2} \frac{\delta S}{\delta \tilde{\eta}^{\mu}_A}|\tilde{g} = \frac{\delta S_{f}}{\delta \tilde{\gamma}^{\mu}_A}.
 \end{eqnarray}

Now one makes the corresponding splits $S_f= S_{f0} + S_{fm}$ and $S=S_0 + S_{ms}.$ The massless parts $S_{f0}$ and $S_0$ go as in the massless case and yield terms that give Einstein's equations and terms that affect only the Rosenfeld total stress-energy tensor.   It remains to consider the equation
$$ \frac{\delta S_{fm}}{\delta \tilde{\gamma}^{\mu}_A}  =  - \frac{ \lambda}{2} \frac{\delta S_{ms} }{\delta \tilde{\eta}^{\mu}_A} | \tilde{g}.$$ 
Assuming the free field mass term to be quadratic in the gravitational potential, and recalling the exclusion of the $C$ term above, 
one can assume the form
$$ S_{fm} = \sqrt{-\eta} \tilde{\gamma}_A^{\mu} \tilde{\gamma}_B^{\nu} ( D \tilde{\eta}_{\mu}^B \tilde{\eta}_{\nu}^A + E \tilde{\eta}_{\mu}^A  \tilde{\eta}_{\nu}^B)_.$$ 
Its variational derivative is
$$ \frac{\delta S_{fm}}{\delta \tilde{\gamma}^{\mu}_A}  = 2 \sqrt{-\eta} \tilde{\gamma}^{\nu}_B (  D \tilde{\eta}_{\mu}^B  \tilde{\eta}_{\nu}^A + E \tilde{\eta}_{\nu}^B  \tilde{\eta}_{\mu}^A)_.$$ Changing to the bimetric variables gives
\begin{equation}
\frac{\delta S_{fm}}{\delta \tilde{\gamma}^{\mu}_A}= \frac{4 \sqrt{-\eta}}{ \lambda}(-D\tilde{\eta}_{\mu}^A  + D \tilde{g}_B^{\nu} \tilde{\eta}_{\mu}^B \tilde{\eta}_{\nu}^A -4E \tilde{\eta}_{\mu}^A + E \tilde{g}_B^{\nu} \tilde{\eta}^B_{\nu} \tilde{\eta}_{\mu}^A). 
\end{equation} 
Thus one infers that 
$$\frac{4 \sqrt{-\eta}}{ \lambda}(-D\tilde{\eta}_{\mu}^A  + D \tilde{g}_B^{\nu} \tilde{\eta}_{\mu}^B \tilde{\eta}_{\nu}^A -4E \tilde{\eta}_{\mu}^A + E \tilde{g}_B^{\nu} \tilde{\eta}^B_{\nu} \tilde{\eta}_{\mu}^A)=  -\frac{ \lambda}{2} \frac{\delta S_{ms} }{\delta \tilde{\eta}^{\mu}_A} | \tilde{g}.$$  One can show that  $$ \frac{ \partial  \sqrt{-\eta} }{ \partial \tilde{\eta}_A^{\mu} }=\frac{1}{4w-1} \sqrt{-\eta} \tilde{\eta}_{\mu}^{A.} $$
For   $S_{ms}$  a natural form to try is 
$S_{ms} = \int d^{4}x (P \tilde{g}^{\nu}_B \tilde{\eta}_{\nu}^B + Q) \sqrt{-\eta}.$
Employing this expression and equating coefficients of like terms determines 
 several of the constants.  Equating the coefficients of the $\sqrt{-\eta} \tilde{\eta}_{\mu}^A$ terms gives $Q = \frac{1}{4 \pi G}(D+4E)(4w-1).$   Equating the coefficients of the $\sqrt{-\eta} \tilde{\eta}_{\mu}^B \tilde{\eta}_{\nu}^A \tilde{g}^{\nu}_B$ terms gives $$P=\frac{D}{4 \pi G}.$$  Equating the coefficients of the 
$\sqrt{-\eta} \tilde{\eta}_{\mu}^A \tilde{\eta}_{\nu}^B \tilde{g}^{\nu}_B$ terms gives  $$P=\frac{E(1-4w)}{4 \pi G}.$$  It follows that $E=\frac{D}{1-4w.}$  
Thus one has 
$$\mathcal{L}_{ms} = \frac{(4w-5)D}{4 \pi G} \sqrt{-\eta} + \frac{D}{4 \pi G} \sqrt{-\eta} \tilde{g}_B^{\nu} \tilde{\eta}^B_{\nu} $$
and $$\mathcal{L}_{fm}= D\sqrt{-\eta} \tilde{\gamma}_A^{\mu} \tilde{\gamma}_B^{\nu} (   \tilde{\eta}_{\mu}^B \tilde{\eta}_{\nu}^A + \frac{1}{1-4w} \tilde{\eta}_{\nu}^B \tilde{\eta}_{\mu}^A).$$

 Combining the algebraic piece of $S_{0}$ with $S_{ms}$ gives 
\begin{eqnarray}
S_{alg}= - \frac{\Lambda}{8 \pi G} \int d^4 x \sqrt{-g} + 2B\int d^4 x \sqrt{ -\eta}  \nonumber \\
+  \int d^4 x\left(  \frac{(4w-5)D\sqrt{ -\eta}}{4 \pi G} + \frac{D\sqrt{ -\eta}}{4 \pi G} \tilde{g}_B^{\nu} \tilde{\eta}_{\nu}^B \right)_.
\end{eqnarray}
When the gravitational potential vanishes, $S_{alg}$ ought to vanish as well.  Imposing this condition to zeroth order gives
$B= \frac{\Lambda  - 8Dw +2D}{16 \pi G}.$
Recalling that the goal is to find a massive generalization of Einstein's theory (with no effective cosmological constant), we require the first-order term in $\tilde{\gamma}^{\mu}_A$ to vanish as well.  It follows that $\Lambda = 2D(4w-1)$ and $B=0.$
   One also expects the quadratic part of the algebraic piece of the action $S_{alg}$ to agree with the free field mass term $S_{fm}.$  After a binomial expansion and some algebra, one sees that this is the case.
One has $$\mathcal{L}_{alg} = D \sqrt{-\eta} \left( \tilde{\gamma}_A^{\mu} \tilde{\eta}_{\mu}^B \tilde{\gamma}_B^{\nu} \tilde{\eta}_{\nu}^A  + \frac{ (\tilde{\gamma}_A^{\mu} \tilde{\eta}_{\mu}^A )^2 }{1-4w}   \right) + HOT,$$
where $HOT$ involves cubic and higher order terms.
   Making a weak-field expansion of the full massive nonlinear action $S$ to relate the coefficient $D$ to the mass $m$ of the spin 2 gravitons shows that $D=-\frac{m^2}{2}.$  For nontachyonic theories, one imposes $D<0.$

	Combining all these results gives the total massive action $S,$ which depends on the spin 2 graviton mass $m$ and the density weight parameter $w$ which controls the relative mass of the spin 0 to that of the spin 2:
\begin{eqnarray}
S= \frac{1}{16 \pi G} \int d^{4}x \sqrt{-g} R(g)  + S_{matter}[\tilde{g}_{\mu\nu}, u]  \nonumber \\ + \frac{1}{2} \int d^{4}x R_{\mu\nu\rho\sigma} (\eta) 
 {\mathcal{M}}^{\mu\nu\rho\sigma}[\tilde{\eta}_{\mu\nu},\tilde{g}_{\mu\nu},u] + \int d^{4}x \partial_{\mu} \alpha^{\mu}       \nonumber   \\
+ \frac{m^2}{8 \pi G} \int d^{4}x ( \sqrt{-g}(4w-1) + \sqrt{-\eta}[5-4w] - \sqrt{- \eta} \, \tilde{g}^{\mu}_A \tilde{\eta}_{\mu}^A),  
\end{eqnarray}
for $w \neq \frac{1}{4}.$

One readily sees that, on fixing the local Lorentz freedom, these theories also fall within the 2-parameter Ogievetsky-Polubarinov family \cite{OP}.  One needs to identify the parameter $w$ with their $-\frac{p}{2}$ and to notice that a tetrad is roughly a  square root of the contravariant tensor $g^{\mu\nu}$,  so their parameter $n$ here takes the value of $\frac{1}{2}.$  One can then use their calculation of the ratio of the spin $0$ mass $m_0$ to the spin $2$ mass $m$:
\begin{equation} m_0 = m \sqrt{ \frac{-16w^2 +24w -5}{8w^2 -12w +4} }_.
\end{equation}
Requiring the mass to be real  leaves two intervals. (By contrast the densitized inverse metric  theories  contribute a    single interval \cite{MassiveGravity1}.)  The mass ratio is shown in the figure. 
\begin{figure}
{\includegraphics[0in,0in][3.in,2in]{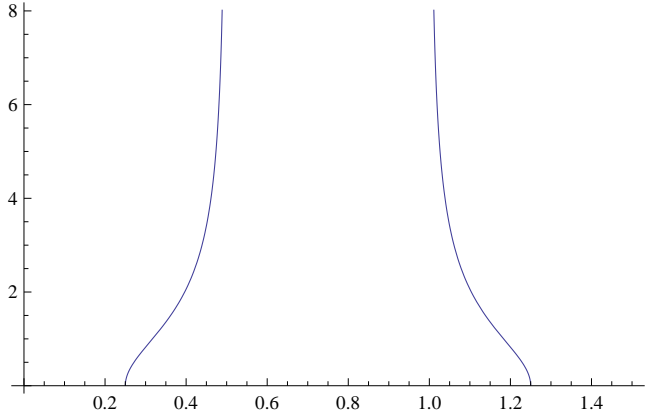}}
\caption{Ratio of Spin $0$ Mass to Spin $2$ Mass as Function of $w$; done with \emph{Mathematica}.}
\end{figure}
 The permitted intervals are $(\frac{1}{4}, \frac{1}{2}]$ and $[1, \frac{5}{4}].$ As $w$ approaches $\frac{1}{4}$ from the right, the spin $0$ mass goes to $0,$ but it never reaches $0$ because $ w \neq  \frac{1}{4}$. The spin 0 mass then rises with $w$; for $w= \frac{3 - \sqrt{3} }{4} \approx .316987,$ $m_0 = m.$  $w=\frac{1}{2}$ gives an infinitely massive spin $0,$ as indicated by the vertical asymptote in the graph, and reduces to the Pauli-Fierz pure spin $2$ case to quadratic order. 
The case $w=1$ on the other branch is similar. The cases $w=\frac{1}{2}$ and $w=1$ correspond, respectively, to the 
third and first of Zumino's pure spin 2 hadron mass terms \cite[pp. 490, 492]{ZuminoDeser}. 
From the vertical asymptote at $w=1$ the mass descends, the ratio reaching $1$ at $w= \frac{3 + \sqrt{3} }{4} \approx 1.18301$. For $w=\frac{5}{4}$  conformal invariance and masslessness for the spin $0$ obtain.


Reconsidering the generalized Bianchi identities for both coordinate and local Lorentz freedom in terms of the bimetric variables, one can readily find the Lorenz-Lorentz-type auxiliary condition.  Using both identities and the gravitational \emph{and} matter field equations, one can infer that 
\begin{eqnarray}
\partial_{\nu} \left( \tilde{\eta}^{\nu}_A  \frac{\delta S }{  \delta \tilde{\eta}^{\mu}_A }|\tilde{g}  -w  \delta_{\mu}^{\nu} \tilde{\eta}^{\alpha}_A   \frac{\delta S }{ \delta \tilde{\eta}^{\alpha}_A }|\tilde{g}     \right) = 0.
\end{eqnarray}
Only the mass term 
$$  \frac{m^2}{8 \pi G}  ( \sqrt{-g}(4w-1) + \sqrt{-\eta}[5-4w] - \sqrt{- \eta} \, \tilde{g}^{\mu}_A \tilde{\eta}_{\mu}^A), $$ gives an interesting result:
\begin{equation}
\partial_{\nu} (\tilde{g}_{B}^{\nu} \tilde{\eta}_{\mu}^B + [1-w] \tilde{g}_B^{\rho} \tilde{\eta}_{\rho}^B \delta_{\mu}^{\nu}) =0. \end{equation}  

Because the mass term is the only place where the tetrads, as opposed to the metrics, appear essentially, it gives an interesting result using the equations of motion, namely, 
\begin{equation}
\tilde{g}^{\rho[B} \tilde{\eta}^{A]}_{\rho} =0, 
\end{equation} 
or equivalently,
\begin{equation}
\tilde{g}^{[\mu}_A \tilde{\eta}^{\nu]A} =0. 
\end{equation}
Thus the potential is symmetric when expressed using indices of the same type.


\section{Conclusion}

Taking into account the results of the present paper and its predecessor \cite{MassiveGravity1}, re-expressing old results using $l=2w$ as needed,  one has these four one-parameter families (with isolated and non-isolated forbidden cases) of universally coupled massive gravities.
For the covariant symmetric tensor (density) potential with weight $-2w,$ one has the universally coupled mass terms  \cite{MassiveGravity1}
\begin{eqnarray}
 \frac{m^2}{16 \pi G}( \sqrt{-g}[1-4w] + \sqrt{-\eta}[4w+1] - \frac{1}{2}\sqrt{- \eta} \, \tilde{g}_{\mu\nu} \tilde{\eta}^{\mu\nu})  
\end{eqnarray}
with   $-\frac{1}{4} \leq w < \frac{1}{4}.$  
For a contravariant symmetric tensor (density) potential with weight $2w,$ one has   \cite{MassiveGravity1}
\begin{eqnarray}
 \frac{m^2}{16 \pi G}( \sqrt{-g}[4w-1] - \sqrt{-\eta}[4w-3] - \frac{1}{2}\sqrt{- \eta} \, \tilde{g}^{\mu\nu} \tilde{\eta}_{\mu\nu}),  
\end{eqnarray}
for $\frac{1}{4} < w \leq \frac{3}{4}.$  
As derived above, for the cotetrad (density) case with weight $-w$ one finds
\begin{eqnarray}
\frac{m^2}{8 \pi G}  ( \sqrt{-g}[1-4w] + \sqrt{-\eta}[4w+3] - \sqrt{- \eta} \, \tilde{g}_{\mu}^A \tilde{\eta}^{\mu}_A).  
\end{eqnarray}
 The permitted intervals for $w$ are $[-\frac{3}{4}, -\frac{1}{2}]$ and $[0, \frac{1}{4}).$ 
Finally, for the tetrad (density) case with weight $w$ one obtains  
\begin{eqnarray}
 \frac{m^2}{8 \pi G}  (\sqrt{-g}[4w-1] + \sqrt{-\eta}[5-4w] -  \sqrt{ -\eta} \tilde{g}_{A}^{\mu} \tilde{\eta}_{\mu}^{A}). 
 \end{eqnarray}
The permitted intervals for $w$ are $(\frac{1}{4}, \frac{1}{2}]$ and $[1, \frac{5}{4}].$ 
 All of them fit within the Ogievetsky-Polubarinov framework, after discarding the local Lorentz freedom if necessary.  The massive tetrad theories, unlike the massive metric theories, permit the spin $0$ mass $m_0$ to be heavier than the spin $2$ mass $m$, not merely lighter or the same, 
 and so have more flexible phenomenology.  
The greater the variety of universally coupled mass terms, the more probable it seems that some of them can escape dangerous resonances sufficiently  to have a chance at stability.  It is therefore important both to pursue maximal generality and to investigate the question of stability, whether analytically, numerically or both.  
In view of the long-range difficulties experienced by General Relativity, such as dark matter and dark energy, there is also empirical motivation for careful exploration of the possibilities for massive gravity.

\section{Acknowledgments}
The author thanks Alexander N. Petrov  and Grant Mathews for discussions and a referee for helpful comments.


\end{document}